\documentclass[
 reprint,
 amsmath,amssymb,
 aps,
 pra,
]{revtex4-2}

\usepackage{graphicx}
\usepackage{dcolumn}
\usepackage{bm}
\usepackage{url}
\usepackage{breakurl}
\usepackage[breaklinks]{hyperref}
\usepackage[T1]{fontenc}
\usepackage[utf8]{inputenc}

\begin{document}

\preprint{APS/123-QED}

\title{Iterative Confinement of Ions via the Quantum Zeno Effect: Probing Paradoxical Energy Consequences}

\author{Varqa Abyaneh}
 \altaffiliation[Also at ]{Opetek, Level 37, 1 Canada Square, Canary Wharf, London, UK}
 \email{varqa.abyaneh@opetek.io}

\date{\today}

\begin{abstract}
Building upon our previously introduced mechanism for ion trapping based on the quantum Zeno effect (QZE), we propose a novel approach to systematically draw ions closer together, solely via quantum measurements. The proposed method involves repeated measurements of the electromagnetic force exerted by ions on an enclosure of conductor plates to confine the ions within an incrementally smaller spatial region, achieved by exploiting the behaviour of the wavefunction at its boundaries. Taking a two-proton system as a case study, we explore the dynamics between the energy gain of the system, attributed to successive QZE measurements, and the energy expended making such measurements. The results reveal a paradox wherein, under specific circumstances, protons appear to accumulate more energy than is seemingly introduced into the system. This peculiarity aligns with prior studies that highlight challenges in energy conservation within quantum mechanics. To verify these observations, we propose an iterative confinement setup that is feasible with current technological capabilities. Confirmation of these findings could offer new insights for applications in quantum physics, including fusion research. Therefore, the proposed novel method of manipulating ions not only harbours considerable potential for diverse applications but also furnishes an additional tool for probing fundamental questions in the field.
\end{abstract}

\maketitle

\section{Introduction}
The Quantum Zeno Effect (QZE), initially proposed by Misra and Sudarshan \citep{misra_zenos_1977}, describes a counterintuitive quantum mechanics phenomenon wherein continuous measurements can effectively `freeze’ the evolution of a quantum system. This phenomenon was later experimentally verified by Itano, Heinzen, Bollinger, and Wineland \citep{QuantumZenoeffect}, who demonstrated that frequent measurements of beryllium ions could inhibit their transition between two distinct energy eigenstates. Subsequent studies have also demonstrated that frequent measurements of a quantum system can constrain it to remain within a specific Hilbert subspace defined by the measurement process \citep{QuantumZenosubspaces}.

Ion trapping, on the other hand, is a technique that aims to confine single or multiple ions into a certain region of space. Achieved through external trapping fields, such as those employed in Paul traps \citep{paul1990electromagnetic}, this technique has a broad spectrum of applications. For example, ion trapping enables the utilisation of ions as qubits for quantum logic operations, thus making quantum computing possible \citep{cirac1995quantum, bernardini2023quantum}. Furthermore, ion traps facilitate high-precision quantum sensing \citep{baumgart2016ultrasensitive} and play integral roles in advanced atomic clocks \citep{ludlow2015optical}, quantum simulators \citep{porras2004effective}, and mass spectrometry \citep{dawson1976quadrupole}. Additionally, they serve as platforms for probing the fundamental laws of physics.

Building on these two areas of quantum mechanics, our previous study \citep{Abyaneh_1} introduced a novel ion trap design based on the QZE, which offers potential advantages over conventional trapping techniques. In particular, unlike traditional methods, our design does not rely on external trapping fields. Therefore, we argued that it can potentially address the challenges associated with traditional methods, such as heating \citep{bruzewicz_trapped-ion_2019}, control \citep{trypogeorgos_synthetic_2018}, and scalability \citep{zhang_versatile_2019}.

Specifically, our previous work \citep{Abyaneh_1} focussed on a one-dimensional model involving two ions within a {QZE confinement region} of length $d$, centred between conductor plates. We demonstrated that the measurements of the force exerted by the electric field of the ions on the conductor plates project the quantum state of the ions onto a Hilbert subspace defined by $d$. Furthermore, measuring this force in an almost continuous manner ensures that the ions remain within the subspace \citep{QuantumZenosubspaces}, thereby effectively creating an ion trap. The study also showed that confining two ions within $10^{-6}$m is feasible with current technology.

Extending our prior work, this study introduces the `iterative confinement' method, a novel approach, never before discussed in the literature, that not only provides ion confinement, but also enables dynamic variations in the size of the confinement region relying solely on quantum measurements. This contrasts with traditional techniques, which require a change in the strength of the confinement field.

This study focuses on using the iterative confinement method to bring two ions closer together despite their Coulomb repulsion, through the use of quantum measurements. The independence of this process from external forces raises a pertinent question regarding the energy dynamics of the system: How does the energy invested in the quantum measurements compare with the resultant change in the ion’s energy?

To better understand this question, it is essential to recognise the complexities that quantum mechanics impose on the principle of energy conservation observed in classical mechanics. These complexities are exemplified by the non-unitary nature of quantum measurements, a topic that has been explored since early contributors such as Wigner \citep{wigner_messung_1952}. 

To provide a specific example illustrating the complexities encountered in energy conservation within quantum mechanics, consider a state $|\psi \rangle$ in the superposition of $n$ energy eigenstates:
\begin{equation}
|\psi \rangle = \sum_{i=0}^n \alpha_i |E_i\rangle.
\end{equation}
The expected energy of $|\psi \rangle$, given the Hamiltonian $\hat{H}$, is:
\begin{equation}
\langle E_{\psi}\rangle \equiv \langle \psi | \hat{H} | \psi  \rangle = \sum_{i=0}^n |\alpha_i|^2 E_i,
\end{equation}
where $E_i$ is the energy eigenvalue of the eigenstate $|E_i\rangle$. Upon successfully measuring the $j^{\text{th}}$ eigenstate and collapsing $|\psi\rangle$ to $|E_j\rangle$, the energy of the quantum system changes to $E_j \neq \langle E_{\psi}\rangle$, seemingly violating energy conservation.

Reconciling the apparent potential for energy non-conservation in quantum theory with the consistent observation of energy conservation in classical mechanics is often discussed in the broader  context of understanding the quantum-to-classical transition. The associated models are typically split into two categories: decoherence models, which do not depend on the collapse postulate, and wavefunction collapse models, which assume an objective physical wavefunction collapse process that is not reliant on an observer.

The implications of energy conservation among these theories are nuanced. For example, the decoherent histories approach \citep{hartle_conservation_1995} and the environmental induced decoherence model \citep{Enviromental_1,joos_emergence_1985,Zurek2007,Environmental_2} eliminate the need for a collapse postulate. They suggested that quantum systems lose coherence due to their interactions with the environment, leading to apparent energy conservation at the macroscopic system level. However, pre-decoherence systems may exhibit unusual energy behaviours owing to incomplete environmental modelling. 

An alternative explanation to the emergence of classical behaviour is the spontaneous localisation concept proposed by Ghirardi, Rimini, and Weber \citep{GRW}, which was later incorporated into the continuous spontaneous localisation (CSL) model \citep{pearle_combining_1989}. This wavefunction collapse model alters standard quantum theory by modifying the Schrödinger equation and introducing an external field. Experimental efforts have been made to constrain the parameters of the CSL model \citep{pearle_bound_1994,pearle_csl_1999,vinante_upper_2016}. However, despite its attempts to explain the quantum-to-classical transition, the model has faced criticism for allowing violations of energy conservation \citep{ballentine_failure_1991,anandan_are_1999}. Pearle \citep{EnergyNonConvservation_2} argued that these concerns are somewhat mitigated when considering the energy of the field causing the collapse.

The potential link between CSL and gravity has been discussed \citep{Pearle_Gravity}, providing the theory with some physical grounding. Other theories have used a more fundamental perspective when exploring the role of gravity in the quantum-to-classical transition. For instance, Penrose \citep{penrose_gravitys_1996,Penrose_2,Penrose_3} hypothesised that the superposition of massive objects leads to the superposition of spacetime geometries, resulting in wavefunction collapse. In contrast, Kay \citep{kay_entropy_1998,kay_decoherence_1998,kay_expectation_2007,abyaneh_robustness_2006} claimed that the quantum state of the gravitational field is unobservable. By tracing over the gravitational degrees of freedom, one can achieve decoherence of the quantum state. However, discussions on whether Penrose’s or Kay’s theories allows for energy non-conservation remain limited.

Other authors have tackled the energy-conservation paradox in quantum mechanics using different approaches. Instead of proposing modifications to the rules of quantum theory, as done in collapse models, or eliminating the measurement collapse postulate like decoherence models do, these authors followed standard quantum theory principles and proposed thought experiments to directly challenge the validity of the principle of energy conservation. Aharonov, Popescu, and Rohrlich \citep{EnergyNonConvservation_3} argue that energy conservation can be violated and that environmental factors alone cannot account for these anomalies. Their hypothesis is rooted in the study of superoscillations, where light behaves beyond the conventional Fourier component limitations.  They propose an experiment to measure these superoscillatory regions. Additionally, Carroll and Lodman \citep{EnergyNonConvservation_1} proposed an experiment predicting measurable energy violations within standard quantum mechanics that could not be explained by environmental factors. Using quantum clocks, Gisin and Cruzeiro \citep{gisin_quantum_2018} deduced that energy is not conserved under certain conssditions. Soltan, Fraczak, Belzig, and Bednorz \citep{soltan_conservation_2021} argued that even in the limit of weak measurements, energy non-conservation is a possibility, suggesting a feasible experiment to verify this. 

This study demonstrates that the QZE-based iterative confinement method can bring ions closer together using less energy than predicted by classical physics. Specifically, we argue that the increase in the energy of a two-proton system resulting from iterative confinement can be greater than the energy expended to perform quantum measurements. While our assumptions regarding energy expenditure are aggressive, experimental verification of the insufficient energy transfer from the photons used to make the quantum measurements, to the two-proton system would also verify this paradoxical result. 

Our findings align with those of previous experiments \citep{EnergyNonConvservation_3,EnergyNonConvservation_1,gisin_quantum_2018,soltan_conservation_2021}. We propose an iterative confinement experiment under achievable practical conditions. The calculated values reflect this feasibility, suggesting that our approach could offer a more effective verification of these claims. Furthermore, our design allows for precise control over experimental parameters, facilitating calibration and future testing. 

An additional advantage of our design is its potential for real-world applications, particularly in fusion research, where efficiently overcoming Coulomb repulsion between ions is critical. While previous experiments aimed to detect potential energy conservation violations, we propose that the iterative confinement method could theoretically repeat this process for multiple iterative steps, thereby amplifying the magnitude of the anomaly.

\section{Results and Discussion} \label{sec:Results}

\subsection{Parameter Choice and Calibration}
This study employs the iterative confinement method outlined in the Methods section (Sec. ~\ref{sec:Iterative Confinement}) for a two-proton system. We adjusted the confinement distance from an initial $d_0 = 10^{-6}\text{m}$ to a final $d_1 = 9.92 \times 10^{-7}\text{m}$, utilising protons with opposite spins to circumvent symmetry constraints arising from the Pauli exclusion principle \footnote{This choice of setup is for illustrative purposes; other setups may prove more practical.}. 

The frequency of QZE confinement measurements was set to $f^{\text{QZE}} = 10^{12}\text{Hz}$. The frequency of the confinement checks at $d_1$ was set to $f^{\text{confine}} = 10^{11}\text{Hz}$. These frequencies were calibrated to be sufficiently small, ensuring that the expected energy cost of making the quantum measurements (equation~\eqref{eq:energy_expected_2}) remained below the associated increase in the two-proton system energy for the iterative step. Additionally, these frequencies align with current technological capabilities.

We proposed a photon frequency $f^{\text{photon}} = 10^7\text{Hz}$ (wavelength $30\text{m}$) for precisely measuring the conductor plate position, chosen to guarantee a sufficiently low expected energy cost of the quantum measurements. However, employing such a wavelength poses a significant challenge, given that a precision of $10^{-10}\text{m}$ may be required \footnote{This precision requirement is based on reasonable assumptions about spring constants. Further details on the practical challenges associated with the QZE-based ion-trapping method are discussed in \citep{Abyaneh_1}.}. However, advancements in interferometry may bridge this 11-order-of-magnitude gap, as evidenced by LIGO's precise measurements using $10^{-6}\text{m}$ wavelength photons capable of measuring displacements of up to $10^{-19}$m \citep{LIGO_abbott_observation_2016}. 

Although the theoretical results presented assume low-frequency photons, which is an impractical assumption, the paradoxical result may be experimentally observable using a more pragmatic frequency such as $10^{14}\text{Hz}$. However, to ensure the integrity of the paradoxical result, it is then imperative to experimentally verify that the measured photons do not significantly transfer energy to the two-proton system.

\subsection{Energy Increase from the Iterative Step}
The system was initially prepared in the ground state of the Hamiltonian $\hat{H}_{d_0}$. Because of this iterative step, the protons are confined to the smaller QZE confinement region $d_1$, thereby increasing the system energy. 

\subsubsection{Quantum Energy Change}
The initial energy of the two-proton system was the ground-state energy.
\begin{equation}
E_{d_0}^{(0)} \equiv \langle \psi_{d_0}^{(0)} |\hat{H}_{d_0} | \psi_{d_0}^{(0)} \rangle.
\end{equation}
Here, $\psi_{d_0}^{(0)}$ denotes the ground state $\hat{H}_{d_0}$. Following successful confinement measurements, the energy of the system is bounded below by the ground-state energy of the Hamiltonian $\hat{H}_{d_1}$. This is because the QZE confinement measurements immediately change, forcing the ions to remain in a superposition of the energy eigenstates of $\hat{H}_{d_1}$. Therefore,
\begin{equation}
\langle E_{d_1} \rangle \geq E_{d_1}^{(0)} \equiv \langle \psi_{d_1}^{(0)} |\hat{H}_{d_1} | \psi_{d_1}^{(0)} \rangle.
\end{equation}
Thus, the quantum energy gain of a two-proton system is bounded as follows:
\begin{equation}
\Delta E_{\text{quantum}}^{\text{ions}} \geq E_{d_1}^{(0)} - E_{d_0}^{(0)}.
\end{equation}
For $d_0=10^{-6}\text{m}$ and $d_1= 9.92 \times 10^{-7}\text{m}$, we obtain
\begin{equation}
\Delta E_{\text{quantum}}^{\text{ions}} \geq 2.63 \times 10^{-24} \text{J}.
\end{equation}
This increase in the energy of the system was purely a result of quantum measurements. This could be verified spectroscopically, which is a challenging yet potentially achievable goal. The probability of a successful confinement measurement is calculated using equation \eqref{eq:confinement_probability} as $17.34$\% \footnote{This probability is a result of our calibration of $d_0$ and $d_1$ to ensure that the predicted outcome is observable.}.

\subsubsection{Classical Energy Change}
From a classical perspective, the change in potential energy owing to the confinement of ions can be approximated as
\begin{equation}
\Delta E_{\text{classical}}^{\text{ions}} \simeq \frac{kq_1^2}{d_1} - \frac{kq_1^2}{d_0}.
\end{equation}
This approximation assumes that the separation between the protons changes from $d_0$ to $d_1$. In our example, substituting the relevant values of $k, q_1, d_0$, and $d_1$ yields
\begin{equation}
\Delta E_{\text{classical}}^{\text{ions}} = 1.86 \times 10^{-24}\text{J}.
\end{equation}
Our quantum energy increase is greater than that of the classical approximation because the iterative confinement step increases the potential and kinetic energies of the protons in the quantum case, whereas the classical calculation is an approximation for the increase in the potential energy of the system. 

\subsection{Expected Energy Expended Via Measurements}
\label{subsec:exp_energy_measurements}
Our theoretical paradox is based on the use of low-energy photons for quantum measurements, and it assumes a complete energy transfer to the two-proton system. Notably, experimental verification showing minimal energy transfer from the measured photons to the protons directly would demonstrate the paradox without relying on the subsequent calculations.

The energy involved in the theoretical model was quantified. Using the formula for the adjusted expected energy expenditure during an iterative step, $\langle E^{\text{QZE}}(d_0,d_1)\rangle^{\text{\text{adj}}}$ (Sec. ~\ref{sec:Survival Probability}, Equation \eqref{eq:adjusted_energy_expected}) and incorporating our variables $f^{\text{photon}}, f^{\text{confine}}, f^{\text{QZE}}$, along with the confinement probability $P(\psi_{d_0},d_0-2\Delta d_0) = 17.34$\%, we find an expected energy expenditure of $7.61 \times 10^{-25}$J.

This value is significantly lower than the lower bound for the quantum energy increase of the system ($2.63 \times 10^{-24}$J) and classical potential energy increase ($1.86 \times 10^{-24}$J). This apparent discrepancy prompts questions regarding energy conservation, suggesting that our understanding of energy dynamics within this quantum measurement framework may be incomplete. 

Our finding that the environment (apparatus) cannot easily account for the energy increase owing to the localisation of protons is supported by \citep{EnergyNonConvservation_1,EnergyNonConvservation_3}, where the authors proposed experiments arguing that predicted changes in energy cannot be explained by the environment.

\subsection{Thermodynamic Implications of the Iterative Method}
At first glance, this result appears to contradict the principles of energy conservation. However, the system under consideration is not closed. The iterative step mandates continuous measurements of the QZE boundaries of $d_0$; upon successful confinement measurements, subsequent continuous measurements must be performed at the new boundary, $d_1$. These continuous measurements imply a continuous input of energy into the system.

Although the open nature of the system offers an explanation, the results remain counterintuitive. This finding has two possible explanations:
\begin{enumerate}
    \item A genuine paradoxical result has been discovered.
    \item The modelling assumptions are inadequate.
\end{enumerate}

\subsubsection{Exploring the Paradoxical Result}
The potential paradox we encountered raises questions regarding the universality of energy conservation in quantum measurements. One intriguing explanation for our results is that the interaction of the photon with the conductor plate, and thus with the electromagnetic field, results in a net energy gain for the ions that exceeds the energy of the photons. This surplus energy may be sourced from the apparatus or the EM field. In a comprehensive quantum state $\psi_{\text{Total}} \in \mathcal{H}_{\text{Total}}$, energy conservation may still hold. Here, we might consider 
\begin{equation}
\label{eq:Total Tensor}
\mathcal{H}_{\text{Total}} = \mathcal{H}_{\text{protons}} \otimes \mathcal{H}_{\text{apparatus}} \otimes \mathcal{H}_{\text{EM-field}} \otimes \mathcal{H}_{\text{other}},
\end{equation}
where $\mathcal{H}_{\text{ions}}$ is the Hilbert space associated with the two-proton system, $\mathcal{H}_{\text{apparatus}}$ is the corresponding Hilbert space for the associated apparatus, and $\mathcal{H}_{\text{EM-field}}$ is the Hilbert space associated with the quantum state of the electromagnetic field owing to the two protons. $\mathcal{H}_{\text{other}}$ may encompass other fields, such as gravity, although such fields are anticipated to be negligible for the scale of our system. 

For energy conservation to hold in the case where we consider the total quantum state $\psi_{\text{Total}}$, the result would still be interesting, suggesting that energy can be transferred between a system and its surroundings in unexpected ways. 

From a practical standpoint, the observed increase in proton energy relative to the energy of the measuring photons could be seen as a transient energy advantage. This advantage persists only until the energy expended to maintain the QZE barrier at $d_1$ exceeds it. In our specific case, the power requirement for maintaining the QZE boundary at $d_1$ was $Q_1^{\text{QZE}} = 1.32 \times 10^{-14}$W. The duration of this energy advantage, $t_{\text{advantage}}$, can be calculated by dividing the energy gain by the power required to maintain the QZE barriers. For our parameters,
\begin{eqnarray}
t_{\text{advantage}} = \frac{\Delta E_{\text{quantum}}^{\text{ions}} - \langle E^{\text{QZE}}(d_0,d_1)\rangle^{\text{\text{adj}}}}{Q_0^{\text{QZE}}} \nonumber\\
 = 1.42 \times 10^{-10}\text{s}.
\end{eqnarray}

\subsubsection{Exploring the Modelling Flaws}
\label{subsec:Limitations}
Note that the paradoxical nature of our results does not necessarily stem from the claim that energy conservation is violated; we have not negated the possibility of energy being sourced from the environment. Furthermore, the nature our results does not stem from an assertion that thermodynamics is being contravened, as our system is not isolated owing to the QZE input energy requirement. Rather, the paradox lies in the observation that the energy of the protons increases by a margin that surpasses the energy of the photons used during the iterative step. 

We should not discount the possibility that the paradoxical result may stem from insufficiently precise modelling of the situation. A more thorough approach might involve modelling the entire system according to quantum field theory, where the photon, conductor plate, and electromagnetic field are treated with greater rigour. Such a detailed analysis may explain the unfeasibility of the experiment, unless a mechanism that transfers the necessary energy from the incoming photons to the system is found.

Furthermore, we acknowledge that our model is based on several assumptions regarding the energy required to make measurements. For instance, we assume that only one photon is required per pulse and that measurements can be conducted with low-frequency photons by leveraging interferometry. While these assumptions do not directly negate our paradoxical result, a field-theoretic approach coupled with these aggressive assumptions could reveal the theoretical unviability of our result.

\section{Conclusion}
Building on the QZE-based ion-trapping method presented in a previous study, we developed an iterative confinement method for the spatial manipulation of ions. Although quantum position measurements involve the localisation of the wavefunction, the iterative confinement method enables us to systematically manipulate the spatial wavefunction over multiple iterative steps, using QZE at each step to prevent undesired wavefunction leakage. 

In particular, we demonstrated how the method can be used to move two ions closer together against their Coulomb repulsion purely via quantum measurements. For a specific setup, the method involves changing the one-dimensional spatial confinement of two protons from a length of $10^{-6}$m to a smaller one of $9.92 \times 10^{-7}$m. Calculations show that the increase in the two-proton system energy is bounded below $2.63 \times 10^{-24}$J. 

However, we argue for the theoretical possibility of creating this energy increase by expending an expected energy of $7.61 \times 10^{-25}$J in quantum measurements, thereby unveiling a paradox. This paradox resonates with the findings of other authors who have conducted various experiments, suggesting that energy dynamics may have to be revisited for quantum systems. 

We argue that one might be able to measure the energy increase in a two-proton system predicted using current technology. However, achieving the predicted low-energy expected expenditure is challenging. Nevertheless, if the energy expended in quantum measurements were experimentally verified not to have been transferred to the two-proton system in sufficient quantities, our paradoxical result would be confirmed. 

An intriguing avenue of exploration is applying iterative confinement to achieve ion fusion. Overcoming the Coulomb barrier to efficiently achieve ion fusion is paramount for the process to have practical value. Our findings demonstrate that, in theory, we can use the iterative confinement method to bring two protons together with less energy than the Coulomb barrier for a single iterative step. Thus, this application may be possible for a setup involving multiple iterative steps and may result in proton fusion. This avenue of research will be discussed in future studies.

\appendix

\section{Methods}
\label{sec:Methods}
The methodologies employed in this study are associated with the UK patent application number GB2314602.0. The complete code and details on reproducing the results of this study are available in the GitHub Repository \href{https://github.com/varqa-abyaneh/Papers/tree/main/Paper_2}{GitHub Repository}. 

\subsection{Proposed Methods}

\subsubsection{Static Confinement}

\begin{figure}[h!]
  \centering
  \includegraphics[width=9.1cm]{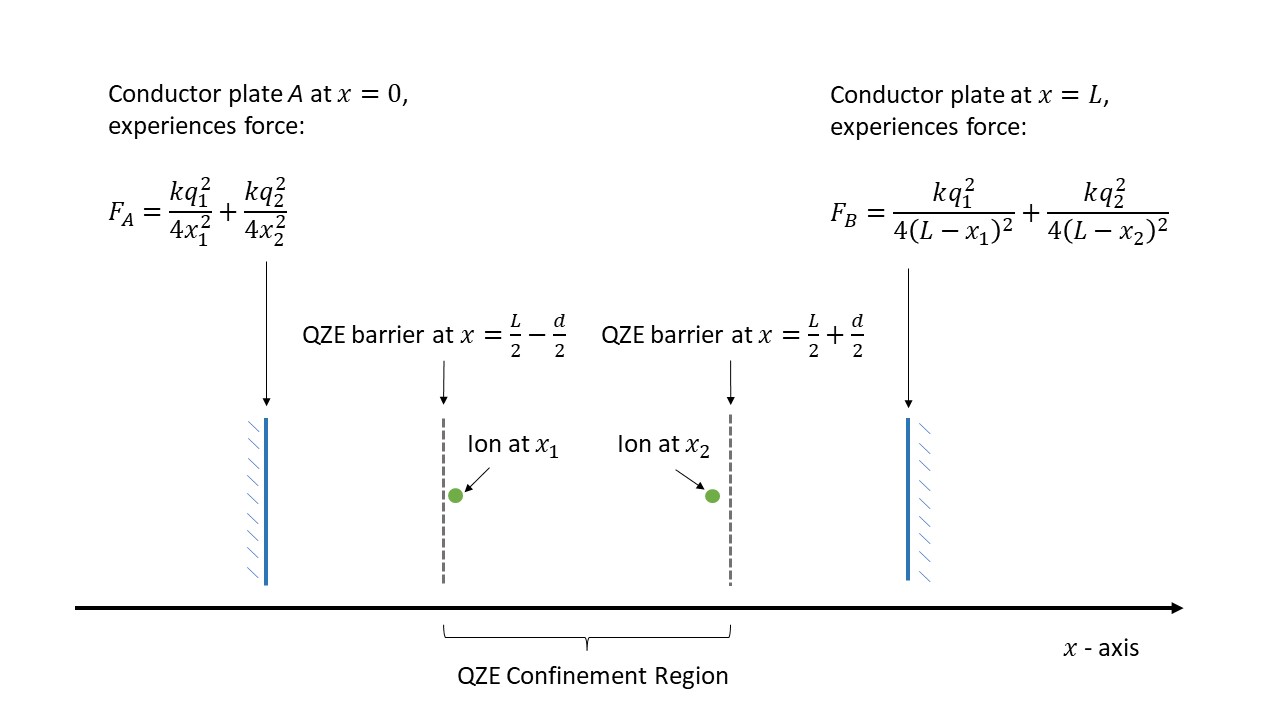}
  \caption{One-dimensional toy model. The conductor plates, set at a distance $L$ apart, measure the force from the two ions. These force measurements indicate if both ions are inside the smaller QZE confinement region of length $d$.}
  \label{Diagram:1-D Toy Model}
\end{figure}

In \citep{Abyaneh_1}, we proposed a static confinement method for confining two ions within a cubic enclosure of edge length $L$, defined by conductor plates, where the approach can be generalised to any number of ions. The central idea revolves around measuring the force from ions on the surrounding conductor plates to infer their position within a smaller \textit{QZE confinement region} with edge length $d<L$. For a simplified one-dimensional representation, as depicted in Fig~\ref{Diagram:1-D Toy Model}, we showed that both ions can be `trapped' in the QZE confinement region provided we make (almost) continuous measurements of the system. These measurements were the total force across the conductor plates, $F_{\text{Total}}$, which must be less than the predetermined force, $F_{\text{Total}}^{\text{Max}}$. $F_{\text{Total}}$ was derived using the method of images in electrostatics; however, the repulsive forces exerted by the conductor plates on each other owing to their induced charges were not considered. This should be considered in a more realistic three-dimensional model. Specifically, we require the following:

\begin{equation}
\label{eq:Projection Constraint}
F_{\text{Total}}<F_{\text{Total}}^{\text{Max}},
\end{equation}
where, 
\begin{eqnarray}
\label{eq:F_Total}
F_{\text{Total}}=F_A+F_B = \frac{kq_1^2}{4x_1^2}+\frac{kq_2^2}{4x_2^2} \nonumber \\ 
              +\frac{kq_1^2}{4(L-x_1)^2}+\frac{kq_2^2}{4(L-x_2)^2}.
\end{eqnarray}
and
\begin{equation}
\label{eq:F_Max}
F_{\text{Total}}^{\text{Max}}=\frac{kq_1^2}{(L-d)^2}+\frac{kq_1^2}{(L+d)^2}+\frac{2kq_2^2}{L^2}.
\end{equation}
Here, the conductor plates are at positions $0$ and $L$ on the $x$-axis, with ions at $x_1$ and $x_2$ and charges $q_1$ and $q_2$; we assume $q_1 \geq q_2$. $k$ is the Coulomb constant in Eqs. \eqref{eq:F_Total} and \eqref{eq:F_Max}. 

\begin{figure}[h]
  \centering
  \includegraphics[width=10cm]{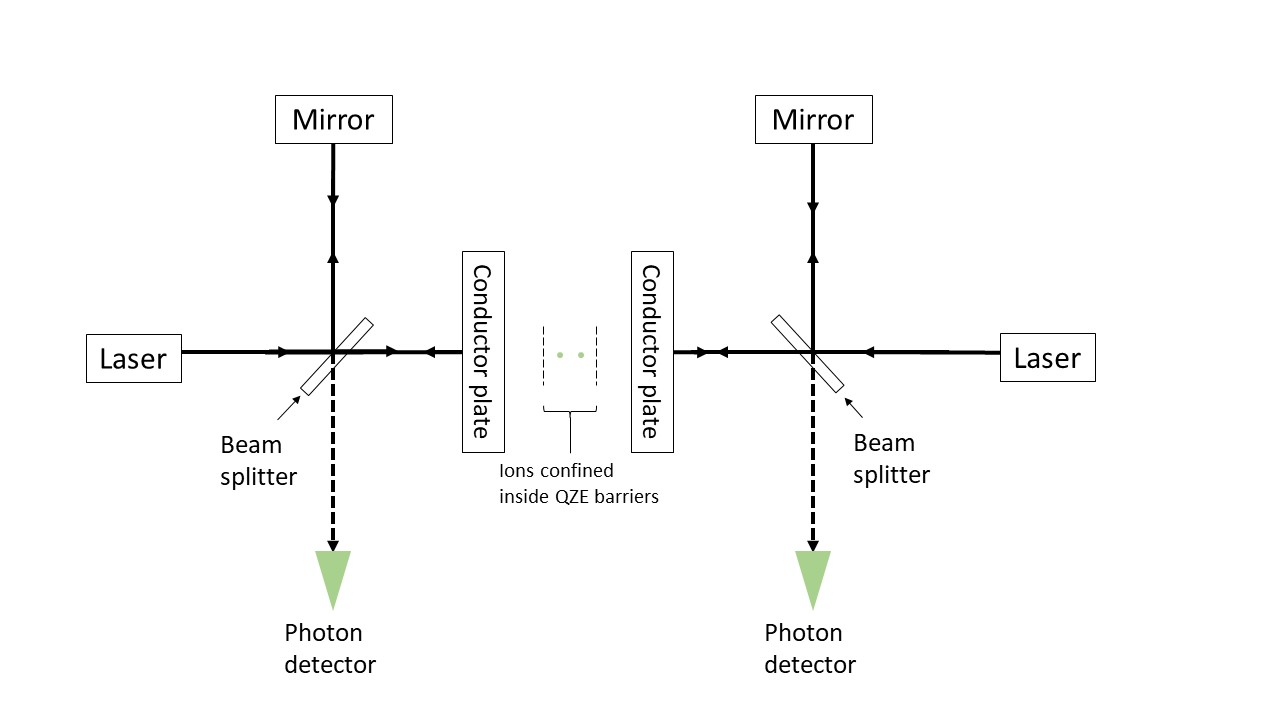}
  \caption{Interferometry setup, whereby lasers can accurately measure the position of the conductor plates, which also act as mirrors to reflect the beam. This enables QZE confinement measurements on the ions.}
  \label{Diagram:Interferometry}
\end{figure}

In \citep{Abyaneh_1}, we proposed that the force on the plates could be deduced by measuring the position of the plate. This can be achieved using interferometry, a technique leveraging light-wave interference \citep{Interferometry}, as shown in Fig~\ref{Diagram:Interferometry} \footnote{Fig~\ref{Diagram:Interferometry} shows two photon detectors, whereas if we were measuring the total force, we would need only one detector. However, an alternative setup is possible, in which the force constraint is on each of the conductor plates.}. Further details of the proposed static confinement method, including practical considerations and experimental challenges, are presented in \citep{Abyaneh_1}.

\subsubsection{Iterative Confinement} \label{sec:Iterative Confinement}

\vspace{2ex} % Adds some space above the title
\noindent (a) \textit{Description.} In the pursuit of iterative confinement, we set the primary objective to establish a one-dimensional QZE confinement region of length $d_0$, symmetrically centred within a larger domain of length $L$ demarcated by the conductor plates. Our methodology involves implementing QZE confinement measurements sustained at a measurement frequency of $f_0^{\text{QZE}}$ to maintain the positions of the ions within $d_0$. Concurrently, we use a separate measurement frequency, $f_0^{\text{confine}}$, to determine whether both ions are localised within a subregion, $d_1=d_0-2\Delta d_0$ ($d_1<d_0$), employing the same technique used for the QZE confinement measurements, except that the measurements check whether the ions are both within the smaller region $d_1$.

Upon successfully confirming ion confinement within $d_1$, we adjusted the QZE confinement to this new region, represented by the measurement frequency, $f_1^{\text{QZE}}$. This process employs measurements at the frequency $f_1^{\text{confine}}$ to verify the presence of ions within a further reduced region $d_2=d_1-2\Delta d_1$ ($d_2<d_1$). This sequence proceeds until the ions are confined within a significantly narrow region, designated as $d_n$. The parameter $\Delta d_i$ represents the reduction in the confinement length at each end for the $i^{\text{th}}$ step, and is set by the experimenter. This methodology facilitates the systematic localisation of ions without directly applying an external force or field, leveraging solely the nuances of quantum measurements. 

\vspace{2ex} % Adds some space above the title
\noindent (b) \textit{Why it works.} \noindent The QZE is central to our method for incrementally confining two ions. The continuous measurement of the force on the conductor plates ensures that the wavefunction of the ions is constrained within the QZE confinement region $d_i$. According to Born's condition on wavefunction continuity, this implies that the wavefunction must be close to zero near the boundaries of $d_i$, thereby facilitating the confinement probability to a slightly smaller region $d_{i+1}$.

\begin{figure}[h]
  \centering
  \includegraphics[width=9cm]{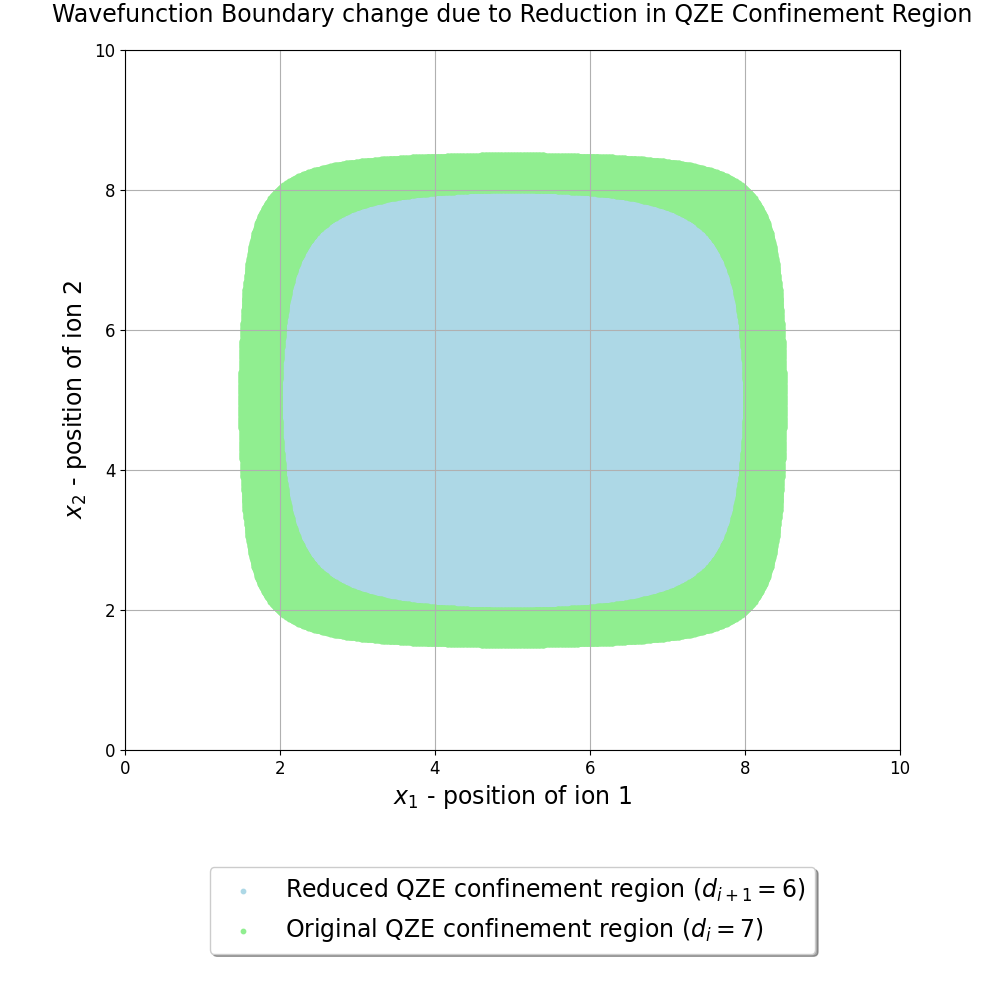}
  \caption{Demonstrating the effect of a successful iterative step on the region where the two-dimensional wavefunction is allowed to be non-zero. $d_i=7$ goes to $d_{i+1}=6$ in equation \eqref{eq:F_Max}, where $k=q_1=q_2=1$, and $L=10$ is assumed for illustrative purposes.}
  \label{Diagram:Wavefunction_Boundaries}
\end{figure}

Fig.~\ref{Diagram:Wavefunction_Boundaries} illustrates this by showing where the wavefunction can be nonzero. Here, the original QZE confinement region is $d_i=7$, and following the confinement measurement, it is reduced to $d_{i+1}=6$. Each axis corresponds to the ion position. The larger green-shaded area is where the wavefunction is allowed to be initially non-zero. 

This is because the QZE confinement measurements force the two-dimensional wavefunction to a region where Equation \eqref{eq:Projection Constraint} holds. $d=d_i=7$ is assumed when calculating $F_{\text{Total}}^{\text{Max}}$ (Equation \eqref{eq:F_Max}) in Equation \eqref{eq:Projection Constraint}. We have also assumed $k=q_1=q_2=1$, $L=10$ in Fig.~\ref{Diagram:Wavefunction_Boundaries} for illustrative purposes. 

The blue-shaded area shows the region where the wavefunction is allowed to be non-zero following a successful confinement measurement. This shaded region is where \eqref{eq:Projection Constraint} is true with $d=d_{i+1}=6$ (all other parameters remain unchanged). The critical point is that $d_i$ and $d_{i+1}$ can be calibrated such that there is a significant probability that the ions will be confined to $d_{i+1}$ if they are already confined to $d_i$. By repeating this iterative step several times, we can confine the ions to increasingly small spaces. 

\subsection{Initial State Preparation}

This study assumes that our initial state is prepared to be in the ground state of the associated confined Hamiltonian. This can be achieved using techniques such as laser or evaporative cooling \citep{teufel_sideband_2011, shuman_laser_2010}. By minimising the thermal fluctuations, we can effectively `force' the system into its lowest-energy eigenstate, establishing a clear starting point for our simulations.

\subsection{Modelling and Numerical Methods} \label{sec:Theoretical Foundations}

\subsubsection{Quantum Zeno Dynamics}

When a single particle undergoes frequent position measurements, which force it into a region of space, the evolution of the particle adds a potential term to the Schrödinger equation \citep{QuantumZenodynamics}. In \citep{Abyaneh_1}, we argued that for a two-ion system, continuous QZE confinement measurements spatially confine the ions to a QZE confinement region defined by \eqref{eq:Projection Constraint}. The quantum Zeno dynamics of this system are argued to be

\begin{align}
\bigg( -\frac{\hbar^2}{2m_1}\frac{\partial^2}{\partial x_1^2} -\frac{\hbar^2}{2m_2}\frac{\partial^2}{\partial x_2^2} + V(x_1,x_2) \bigg. \notag \\
\bigg.  +U(x_1,x_2) \bigg) \psi(x_1,x_2)=E\psi(x_1,x_2).
\label{eq:QZE_SE}
\end{align}
where $m_1$ and $m_2$ are the masses of the ions, $\psi(x_1, x_2)$ is the wavefunction of the system depending on the positions $x_1$ and $x_2$ of the two ions, and $V(x_1, x_2)$ is the potential energy function arising from the repulsive Coulomb interaction between the ions. $U(x_1, x_2)$ is the `effective' potential energy function arising from the QZE measurements, which can be conceptualised as follows:
\begin{equation}   
\label{eq:U_QZE} 
U(x_1, x_2) = 
    \begin{cases}
    0 \text{     if  } F_{\text{Total}} < F_{\text{Total}}^{\text{Max}} \\
    \infty \text{    otherwise}
    \end{cases}.
\end{equation}
Here, $F_{\text{Total}}$ and $F_{\text{Total}}^{\text{Max}}$ are given by Equations \eqref{eq:F_Total} and \eqref{eq:F_Max}, respectively. 

\subsubsection{PDE Grid and Time Evolution}

A grid-based finite-difference technique is employed to calculate the eigenstates of Equation \eqref{eq:QZE_SE}. To avoid singularities in the Coulomb potential $V(x_1, x_2)$ when $x_1=x_2$, a regularisation term $\epsilon$ is employed, as follows:

\begin{equation}
V(x_1,x_2) = k \frac{q_1q_2}{\sqrt{(x_1-x_2)^2+\epsilon^2}}.
\end{equation}
$\epsilon$ is set to $10^{-15}$m and justified by noting that when the denominator of the above Coulomb potential is below this number, we expect a strong nuclear force to dominate \citep{wong1998introductory}.

The Crank--Nicolson scheme is a numerical method used for solving time-dependent differential equations, offering stability and accuracy for evolving wavefunctions \citep{NumericalRecipes}. 

\begin{multline}
\left( \hat{I}-\frac{i\Delta t}{2\hbar} \hat{H} \right) \psi(i,j,k+1) \notag \\
   = \left( \hat{I}+\frac{i\Delta t}{2\hbar}\hat{H} \right)\psi(i,j,k),
\end{multline}
where $\hat{I}$ is the identity matrix, and $\hat{H}$ is the Hamiltonian matrix \footnote{Here, the Hamiltonian does not include the effective potential term $U(x_1,x_2)$ because we are evolving the state between successive QZE confinement measurements.}. $\psi(i,j,k)$ is the state with $i,j$ representing the spatial grid point of each ion, $k$ is the time step, and $\Delta t$ is the size of the time step.

\subsubsection{Leakage Function}

In \citep{Abyaneh_1}, the \textit{ leakage function} $L(\psi_{d},t_{QZE})$ was introduced to serve as a measure of the extent to which the wavefunction extends beyond a predefined boundary between successive QZE confinement measurements $t_{\text{QZE}}$,
\begin{equation}
\label{eq:leakage_function}
L(\psi_{d},t_{QZE}) = \int\limits_{\text{outside}} |\psi_{d}(x_1, x_2, t_{QZE})|^2 \, dx_1 \, dx_2.
\end{equation}
Here, $\psi_{d}(x_1, x_2, t_{\text{QZE}})$ represents the wavefunction in two dimensions at $t_{\text{QZE}}$ after the wavefunction was previously trapped within the region defined by $d$. The integral is calculated over the region outside the region defined by $d$, specifically in areas where Equation \eqref{eq:Projection Constraint} is not satisfied.

\subsubsection{Confinement Probability} \label{sec:Prob_Confinement_Fusion}

We introduce the \textit{confinement probability} function, which is pivotal for analysing the iterative confinement method. We assume that at each step, the ions are in the ground state of the Hamiltonian $\hat{H}_{d_i}$, which can certainly be achieved in the first step, $i=0$. Assuming this, we calculated the probability that the ions are confined within a reduced distance $d_{i+1}=d_i-2\Delta d_i$ as follows:

\begin{equation}
\label{eq:confinement_probability}
P(\psi_{d_i},d_i-2\Delta d_i) = \int\limits_{\substack{\text{inside} \\ d_i-2\Delta d_i}} |\psi_{d_i}(x_1, x_2)|^2 \, dx_1 \, dx_2.
\end{equation}
Here, $\psi_{d_i}(x_1,x_2)$ is the ground-state wavefunction of the Hamiltonian $\hat{H}_{d_i}$. This probability is critical for deriving energy expenditure in the iterative setup. The integral is obtained over the region where both ions are confined within $d_i - 2 \Delta d_i$. This is the region where equation \eqref{eq:Projection Constraint} is true, and $d = d_i - 2\Delta d_i$ is set in equation \eqref{eq:F_Max}.

\subsubsection{Expected Time for Iterative Step}\label{sec:Expected Time}

The computation of the expected time for successful measurement, confirming the ions within the more confined region $d_{i+1} < d_i$, is predicted based on a critical assumption, denoted as Assumption A. \\

\noindent \textbf{Assumption A.} \textit{Should the measurement aimed at confining both ions within $d_{i+1}$ fail, any subsequent measurement retains an equal or greater likelihood of success.} \\

In essence, Assumption A implies modelling a failed confinement measurement as `non-intrusive'. Our results did not depend on a successful measurement that elevated the system to an energy state higher than the ground state of $\hat{H}_{d_{i+1}}$. When considering multiple iterative steps, we conservatively adopt the approachassume that a successful confinement measurement changes the stateleads to the ground state of the associated Hamiltonian.

This nuanced understanding of the influence of measurement is vital for the accuracy of our time estimations and integrity of our expected energy expenditure calculations. The expected time for an iterative step for the ions to be confined to $d_{i+1}$, since being QZE confined to $d_i$, is given by

\begin{equation}
\label{eq:t_expected}
\langle t(\psi_{d_i}, d_i-2\Delta d_i)\rangle = \frac{1}{f_i^{\text{confine}} P(\psi_{d_i}, d_i-2\Delta d_i)},
\end{equation}
where $P(\psi_{d_i}, d_i-2\Delta d_i)$ denotes the probability defined in equation \eqref{eq:confinement_probability}. This equation has several justifications.

\begin{itemize}
\item \textbf{Physical Interpretation}: A shorter expected time arises with more frequent measurements or a higher probability of the ions in the desired region.
\item \textbf{Dimensional Consistency}: This equation is dimensionally consistent.
\item \textbf{Limiting Cases}: The equation behaves as anticipated in extreme cases, demonstrating is accuracy.
\end{itemize}

\subsubsection{Expected Energy for Iterative Step}\label{sec:Expected Energy}
The expected energy expended via the iterative confinement method depends on the expected time $\langle t(\psi_{d_i}, d_i-2\Delta d_i) \rangle$ for a successful measurement that confines the ions within the region $d_i - 2\Delta d_i$. The expected energy for each iterative step is the product of the QZE confinement power $Q_i^{\text{QZE}}$ and associated expected time.

\begin{equation}
\label{eq:energy_expected_1}
\langle E^{\text{QZE}}(\psi_{d_i}, d_i-2\Delta d_i)\rangle = Q_i^{\text{QZE}}\langle t(\psi_{d_i}, d_i-2\Delta d_i)\rangle.
\end{equation}
Considering the energy of a photon as $hf^{\text{photon}}$, the power required for maintaining the QZE boundaries is

\begin{equation}
\label{eq:Q_QZE}
Q_i^{\text{QZE}}=2hf^{\text{photon}}f_i^{\text{QZE}}.
\end{equation}
The factor of two in the above equation comes from the fact that we have a QZE boundary at each end of $d_i$ \footnote{This factor of two will be higher in a more realistic three-dimensional setup. For example, a cubic enclosure of six conductor plates would have a factor of six.}. Assuming that only one photon per pulse is required to make a successful measurement of the QZE boundary is ambitious. In reality, we expect this number to be larger because we consider real-world factors, such as detector sensitivity. By substituting equations \eqref{eq:t_expected} and \eqref{eq:Q_QZE} into equation \eqref{eq:energy_expected_1}, we obtain

\begin{multline}
\label{eq:energy_expected_2}
\langle E^{\text{QZE}}(\psi_{d_i},d_i - 2 \Delta d_i)\rangle  \\
=\frac{2hf_i^{\text{QZE}}f^{\text{photon}}}{f_i^{\text{confine}}P(\psi_{d_i},d_i-2\Delta d_i)}.
\end{multline}
Given the nuances of the iterative confinement method, the overlap between $f_i^{\text{QZE}}$ and $f_i^{\text{confine}}$ is essential.

\subsubsection{Survival Probability and Adjusted Expected Energy}\label{sec:Survival Probability}

The expected energy, as derived in Equation \eqref{eq:energy_expected_2}, assumes that the wavefunction remains intact with no leakage between successive QZE confinement measurements. In \citep{Abyaneh_1}, we defined what would be done in the event of a failed QZE confinement measurement. Here, we assume a more conservative course of action so that our calculation of the expected energy expended on moving the ions from the ground state of $\hat{H}_{d_0}$ to the ground state of $\hat{H}_{d_n}$ is also conservative. This was achieved by assuming that the entire experiment must be restarted in the event of a failed QZE confinement measurement. We can approximate the survival probability $P^{\text{survival}}$ for the iterative confinement method with $n$ steps as follows:

\begin{multline}
P_i^{\text{survival}} =  \\
 \left( 1 - L(\psi_{d_i},t_i^{\text{QZE}})\right)^{f_i^{\text{QZE}} \langle t(d_i,d_i - 2 \Delta d_i)\rangle}.
\end{multline}
In this context, the expected number of measurements at step $i$ is derived by multiplying the QZE confinement measurement frequency by the expected time associated with that step. A key assumption is the independence of probabilities across steps. This is equivalent to the assertion that no measurement perturbs the quantum state. The survival probability for the entire experiment over multiple $i$ steps can then be expressed as the product across all $i$ steps:

\begin{equation}
\label{eq:survival_probability}
P^{\text{survival}}(d_0,d_n) = \prod_{i=0}^{n-1} P_i^{\text{survival}}.
\end{equation}
We can now adjust the expected energy calculation in Equation \eqref{eq:energy_expected_2} to account for the survival probability and provide a more meaningful quantity:
\begin{equation}
\label{eq:adjusted_energy_expected}
\langle E^{\text{QZE}}(d_0,d_n)\rangle^{\text{\text{adj}}} = \frac{\langle E^{\text{QZE}}(d_0,d_n) \rangle }{P^{\text{survival}}(d_0,d_n)}.
\end{equation}
We argue that this adjusted expected energy may be seen as a more practical measure, given that it also models the possibility of experimental failure at any step $i$ because of wavefunction leakage \footnote{Note that we assume zero energy expenditure in resetting an experiment following a failed confinement. While unrealistic, we argue that this does not detrimentally affect the theoretical validity of our results.}. While this derivation contains numerous assumptions, we believe that many of them are theoretically conservative. This means that performing the iterative confinement method would be theoretically cheaper, were we to enhance the model. As discussed in Sec.~\ref{subsec:Limitations}, while our model is conservative in its theoretical assumptions, this is unlikely to translate into practical conservativeness in experimental settings. 

\section*{Acknowledgements}
I am grateful to Luke Hopkins and Philip Beadling for their help and advice on the Python code. I would further like to thank Editage (www.editage.com) for English language editing.

\bibliography{references_paper_2}

\end{document}